\documentclass[prb,amsfonts,twocolumn,amssymb,floats,aps]{revtex4}

\usepackage[final,dvips]{epsfig}
\usepackage{bm}

\begin{document} 

\title[Correlation effects in APS of Ni]
{Electron-correlation effects in appearance-potential spectra of Ni} 

\author{M. Potthoff}

\email[email: ]{michael.potthoff@physik.hu-berlin.de}

\author{T. Wegner}
\author{W. Nolting}

\affiliation{
Lehrstuhl Festk\"orpertheorie,
Institut f\"ur Physik, 
Humboldt-Universit\"at zu Berlin, 
10115 Berlin, Germany
}

\author{T. Schlath\"olter}

\affiliation{
Division Technical Systems, Philips Research, 22335 Hamburg, Germany
}

\author{M. Vonbank}
\author{K. Ertl}

\affiliation{
Max-Planck-Institut f\"ur Plasmaphysik, EURATOM Association, 
85740 Garching bei M\"unchen, Germany
}

\author{J. Braun}
\author{M. Donath}

\affiliation{
Physikalisches Institut,
Universit\"at M\"unster,
48149 M\"unster, Germany
}

\begin{abstract}
Spin-resolved and temperature-dependent appearance-potential spectra
of ferromagnetic Nickel are measured and analyzed theoretically.
The Lander self-convolution model which relates the line shape 
to the unoccupied part of the local density of states turns out 
to be insufficient. Electron correlations and orbitally resolved
transition-matrix elements are shown to be essential for a 
quantitative agreement between experiment and theory.
\end{abstract}


\maketitle 

\section{Introduction} 

Due to the Coulomb interaction, an electron approaching a metal surface 
can excite a core electron into a state of the unoccupied bands. 
Above a threshold energy both, the excited core electron and the de-excited 
primary electron are scattered into valence states just above the Fermi 
energy. The appearance-potential spectroscopy (APS) 
\cite{PH74,WE76,DR83,PER92} monitors the intensity 
of this radiationless transition as a function of the energy of the 
primary electrons by detecting the emitted X-rays or Auger electrons of 
the subsequent core-hole decay. 
Since core-hole formation is involved, the method is element specific and 
local.
Its comparatively simple experimental setup and its surface sensitivity 
qualifies APS as a useful technique for surface analysis.
For a ferromagnetic material, the spin dependence of the AP signal obtained 
by using a polarized electron beam reflects the surface 
magnetization as has been demonstrated for the transition metals Fe and Ni.
\cite{Kir84,EVDN93,DVDD95,RKRD00,KRRD00}
In a sense APS can be considered to be complementary to Auger-electron 
spectroscopy (AES). \cite{Ram91}
While the Auger line shape resulting from core-valence-valence (CVV) 
transitions yields information on the occupied part of the valence band, 
APS is sensitive to the unoccupied electronic structure.
Contrary to $\bf k$ resolved (inverse) photoemission, the AP transition is 
more or less localized in real space.
This suggests to employ APS as a quantitative probe for the unoccupied part 
of the (spin-dependent) local valence density of states (DOS).

In most cases the interpretation of appearance-potential spectra is still 
based on a simple independent-electron model suggested by Lander in 1953: 
\cite{Lan53} 
Hereafter, the line shape is given by the self-convolution of the unoccupied 
part of the DOS. 
The desired information can then be obtained by de-convolution techniques.
\cite{DF79,SSRC84}
Within the context of AES, however, it soon became clear that the 
self-convolution model seems to be oversimplified. 
Powell \cite{Pow73} discovered an ``anomalous'' shape of the CVV Auger line 
of Ag which has been attributed to electron-correlation effects. 
Correlations may be significant also for APS because of the {\em direct} 
interaction of the two additional valence electrons 
in the final state. 
This is demonstrated by the Cini-Sawatzky theory: \cite{Cin77,Saw77}
Within the framework of the single-band Hubbard model, the two-particle 
(APS/AES) excitation spectrum is dominated by a pronounced satellite feature 
which is split off the band-like part by a characteristic energy of the order 
of the on-site Coulomb interaction $U$. 
For more realistic (multi-band) Hubbard models the {\em direct}
correlations give rise to very complex satellite structures.
\cite{KD89,NGE92,PBNB93,PBB94,PBBN95}

A second shortcoming of the Lander model consists in the fact that
transition-matrix elements are not taken into account.
Modern theories for APS (AES) based on density-functional theory and 
the local-density approximation (LDA) 
have overcome this deficiency. \cite{HWMR88,EP97}
Although the direct correlations are still neglected in these approaches, 
it seems that the AP spectra of Fe and Ni, for example, are well understood
\cite{EP97,RPP97} --
serious indications for correlation effects have not been observed.
This is surprising since it is well known that the {\em Auger} line shape 
of Ni cannot be explained within an effective independent-electron model.
\cite{TDDS81}
It is dominated by a strong spectral-weight transfer due to an on-site 
interaction of the order of $U\approx 2-3$ eV to be compared with the 
effective $d$-band width $\Delta \approx 3$~eV. \cite{Ram91,BFHLS83}

Clearly, the question what really determines the appearance-potential 
line shape of transition 
metals, can only be answered {\em a posteriori} -- namely by comparison with 
a theory that realistically includes electron-correlation as well as 
matrix-element effects, orbital degeneracy and $sp$-$d$ hybridization from 
the very beginning. 
The analysis of the AP line shape of Ni, as a prototypical ferromagnetic 
$3d$ metal, is the purpose of the present paper.
In the experiment spin-resolved spectra from excitation of the 
$2p_{3/2}$ ($L_{\rm{III}}$) core level are recorded for different 
temperatures ranging from $T=100$~K up to and slightly above the Ni Curie 
point $T_{\rm C}=624$~K. \cite{LB19a}
The theoretical interpretation of the line shape is essentially based on 
a multi-band Hubbard-type model with general on-site Coulomb interaction 
and a realistic parameterization of the (LDA) one-particle electronic 
structure as an input for a subsequent many-body calculation. 
Standard diagrammatic techniques are used to account for correlations. 
The transition-matrix elements are calculated within the usual intra-atomic 
approximation.
The novel feature of the approach is that it allows to study {\em separately} 
the DOS effect (bare self-convolution), the temperature-dependent effect of 
the (direct and remaining indirect) correlations as well as the effect of the 
transition-matrix elements.

\section{Experiment} 

Experiments are performed for a Ni(110) single-crystal surface with 
in-plane magnetization. \cite{Don89,EVDN93}
The spin-polarized electron beam used for excitation is emitted 
from a GaAs photocathode irradiated by circularly polarized light. 
The longitudinal spin polarization of the emitted photoelectrons is changed 
into transversal by $90^\circ$ electrostatic deflection.
To correct for the incomplete polarization of the electrons ($P \approx 30\%$),
all data have been rescaled to a $100\%$ hypothetical beam polarization
and alignment between the electron-polarization and the sample-magnetization 
vector.
In the present setup the core-hole decay is detected via soft-X-ray emission 
(SXAPS). 
The detector arrangement consists of a multichannel plate with filters and 
a CsI layer acting as photon-to-electron converter.
The APS signal is measured as a function of the primary energy of the 
electrons.
Modulation of the sample potential together with lock-in technique is 
employed to separate the signal from the otherwise overwhelming background.
For the potential modulation a peak-to-peak voltage of $2V$ has been chosen.
This value ensures high APS signals for the $2p_{3/2}$ core level without 
modulation-induced broadening effects.
Further details of the experimental setup and the sample preparation have 
been described elsewhere. \cite{EVDN93}

\begin{figure}[t]
\begin{center}
\epsfig{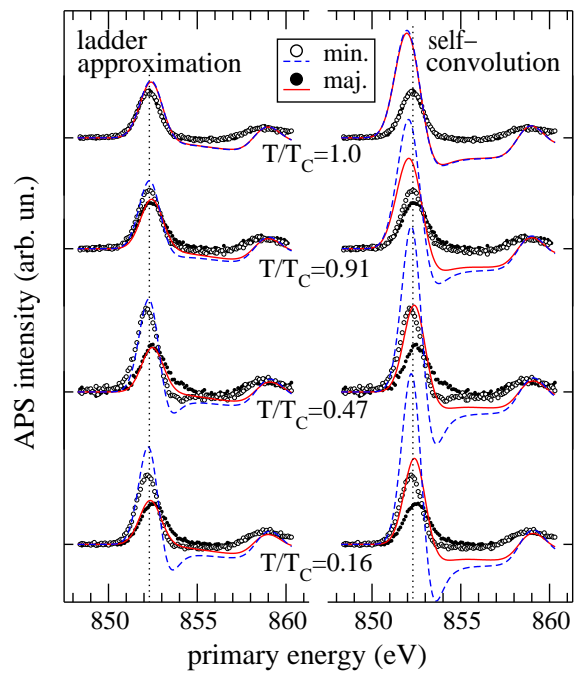}

\caption{
Spin-resolved $L_{\rm III}$ Ni appearance-potential spectrum for 
different temperatures $T/T_{\rm C}$.
{\em Data points:} measured differential intensity $dI/dE$ as a 
function of the primary energy. For better comparison with 
theory the same data are shown twice (left and right panel).
{\em Lines, left:} 
theory with direct and indirect correlations included 
(ladder approximation).
{\em Lines, right:} indirect correlations included only 
(self-convolution).
\label{fig:spectrum}
} 
\end{center}
\end{figure}

Fig.\ \ref{fig:spectrum} shows the differential AP intensity as a function 
of the energy of primary electrons with polarization parallel (minority, 
$\downarrow$) or 
antiparallel (majority, $\uparrow$) to the target magnetization.
The displayed energy range covers the emission from the $L_{\rm III}$
transition. 
The $L_{\rm II}$ emission would be visible at higher energies shifted by 
the $2p$ spin-orbit splitting of about 17.2~eV.

For $T/T_{\rm C}=0.16$ the system is close to ferromagnetic saturation.
The appearance-potential spectrum shows a strongly asymmetric intensity 
ratio as well as a spin splitting of the main peak at $E=852.3$~eV 
(dotted line).
Since Ni is a strong ferromagnet, there are only few unoccupied $d$ states 
available in the majority spin channel, and thus $I_\uparrow < I_\downarrow$ 
holds for the (non-differential) intensities. 
This is the dominant spin effect. 
The intensity asymmetry in the main peak gradually diminishes with 
increasing temperature and vanishes at $T_{\rm C}$.
Note that only the position of the majority peak shifts with $T$ while 
the minority peak position remains unchanged.

The main peak is related to the high density of states at the Fermi energy 
and is mainly due to $d$-$d$ character of the two final-state electrons. 
Additional small $s$-$d$ contributions are present in the secondary peak 
at $E \approx 859$~eV as has been concluded from the analysis of the 
transition-matrix elements. \cite{EVDN93,EP97}
This structure has been identified as originating from a DOS discontinuity 
deriving from the $L_7$ critical point in the Brillouin zone. \cite{EVDN93}
No temperature dependence and spin asymmetry is detectable here.

\section{Theory}

The theoretical approach is based on the usual two-step description assuming 
the lifetime of the core hole to be sufficiently long. \cite{PH74,Cin79}
The AP line shape is then unaffected by the cross section for the core-hole 
decay step but solely determined by the excitation step.
According to Fermi's golden rule, the intensity can be written as:
\begin{eqnarray}
  I_{\sigma_c\sigma_i} ({\bf k}_\|,E) & \propto &
  \mbox{Im} 
  \sum_{L_1L_2L_1'L_2'} 
  M_{L_1L_2}^{\sigma_c\sigma_i}({\bf k}_\|,E) \times
\nonumber \\ && \mbox{} \hspace{-22mm}
  \langle \langle 
    c_{iL_1\sigma_c} 
    c_{iL_2\sigma_i} ; 
    c_{iL_2'\sigma_i}^\dagger 
    c_{iL_1'\sigma_c}^\dagger 
  \rangle \rangle_E \;
  ( M_{L_1'L_2'}^{\sigma_c\sigma_i}({\bf k}_\|,E) )^\ast
  \: .
\label{eq:int}
\end{eqnarray}
For an incoming electron with spin $\sigma_i=\uparrow,\downarrow$, 
one has to consider the ``singlet'' transition, i.~e.\ excitation 
of a core electron with spin $\sigma_c = -\sigma_i$, as well as the 
``triplet'' transition with $\sigma_c = \sigma_i$. 
Since the spin state of the final core hole is not detected, the 
intensities have to be summed incoherently:
\begin{equation}
  I_{\sigma_i}({\bf k}_\|,E) \equiv 
  I_{\sigma_c\sigma_i}({\bf k}_\|,E) + I_{-\sigma_c\sigma_i}({\bf k}_\|,E) 
  \: . 
\end{equation}
Above the Curie temperature $I_{\sigma_i} = I_{-\sigma_i}$. 

The ``raw spectrum'' resulting from an intra-atomic transition at 
the lattice site $i$ is given by the imaginary part of the retarded
two-particle (Zubarev \cite{Zub60}) Green function in Eq.\ (\ref{eq:int}).
It describes the correlated propagation of the two additional final-state 
electrons at energy $E$:
\begin{eqnarray}
  && \langle \langle 
    c_{iL_1\sigma_c} 
    c_{iL_2\sigma_i} ; 
    c_{iL_2'\sigma_i}^\dagger 
    c_{iL_1'\sigma_c}^\dagger 
  \rangle \rangle_E =
  \nonumber \\ &&
  - \int_0^\infty dt \: e^{iEt} \langle [
    c_{iL_1\sigma_c} (t)
    c_{iL_2\sigma_i} (t) ; 
    c_{iL_2'\sigma_i}^\dagger (0)
    c_{iL_1'\sigma_c}^\dagger (0)
    ]_- \rangle \: .
    \nonumber \\
\end{eqnarray}
Here $\langle \cdots \rangle$ is the thermodynamical average, $[...,...]_-$ 
the commutator, and ${\cal O}(t) = \exp(iHt){\cal O}\exp(-iHt)$ is the
Heisenberg time dependence of an operator ${\cal O}$.
The Green function is expressed in terms of annihilators (creators) 
$c^{(\dagger)}_{iL\sigma}$ which refer to a tight-binding one-particle
basis. 
Due to translational symmetry of the fcc lattice the $i$ dependence is 
only formal. 
$L=\{l,m\}$ characterizes the angular momentum of the localized $3d$, $4s$, 
and $4p$ basis functions.
In total there are $2 \times 9$ orbitals 
$| i L \sigma \rangle = c_{iL\sigma}^\dagger | \mbox{vac.} \rangle$
per site.
 
Because of the definite $L$ character of the basis orbitals, the orbitals 
on different sites are mutually non-orthogonal:
$\langle iL\sigma | i'L'\sigma' \rangle = S_{ii'}^{LL'} \delta_{\sigma\sigma'}$.
There are different advantages to develop the many-body formalism for a 
non-trivial overlap matrix ${\bf S} \ne {\bm 1}$ (see Refs.\ 
\onlinecite{DDVP93,WPN00}):
A unique decomposition of the spectrum into $s$, $p$, and $d$ parts is
possible, the basis orbitals are more localized, \cite{DDVP93} and in the 
interaction part of the Hamiltonian the on-site Coulomb interaction among 
the $3d$ electrons can be dealt with separately. 
Furthermore, the same basis states enter the definition of the 
transition-matrix elements, and their definite angular-momentum 
character implies helpful selection rules.

The transition-matrix elements in Eq.\ (\ref{eq:int}),
\begin{eqnarray}
  && M_{L_1L_2}^{\sigma_c\sigma_i}({\bf k}_\|,E) = \nonumber \\ &&
  {}^{(1)} \langle \mbox{2p}, \sigma_c |           \:
  {}^{(2)} \langle {\bf k}_\| E \sigma_i |         \:
  \frac{1}{|{\bf r}_1-{\bf r}_2|}                 \:
  | i L_1 \sigma_c \rangle^{(1)}                \:
  | i L_2 \sigma_i \rangle^{(2)} \: , \nonumber \\
\label{eq:tme}
\end{eqnarray}
are calculated by assuming the transition to be intra-atomic as usual.
\cite{Ram91,HWMR88,EP97}
The different wave functions as well as the Coulomb operator 
$1/|{\bf r}_1-{\bf r}_2|$ are 
expanded into spherical harmonics, the angular integrations are performed 
analytically, and the numerical radial integrations are cut at the 
Wigner-Seitz radius (see Ref.\ \onlinecite{Sch00} for details).

Surface effects enter the theory via the high-energy scattering state 
$|{\bf k}_\|E\sigma_i\rangle$.
It is calculated as a conventional LEED state \cite{Pen74} with ${\bf k}_\|=0$ 
to describe the normally incident electron beam in the experimental setup.
At kinetic energies of the order of keV, however, the ${\bf k}_\|$ dependence 
turns out to be weak. 
Furthermore, at high kinetic energies, multiple-scattering effects are small 
and may be neglected for convenience.

The (paramagnetic) LDA potential for Ni is determined by a self-consistent 
tight-binding linear muffin-tin orbitals (LMTO) calculation. \cite{AJ84}
The $3d$, $4s$, and $4p$ valence orbitals 
$| i L \sigma \rangle$ are taken 
to be the muffin-tin orbitals (MTO's).
The four-fold degenerate $2p_{3/2}$ core state is obtained from the LDA core 
potential by solving the radial Dirac equation numerically. 
Its (relativistically) large component is decomposed into a (coherent) 
sum of Pauli spinors $| \mbox{2p}, \sigma_c \rangle$ with 
$\sigma_c = \uparrow, \downarrow$.

Note that the Green function in Eq.\ (\ref{eq:int}) generally
depends on {\em four} quantum numbers $L_{1}$--$L_{4}$. In fact, 
for the present case each term in the sum gives a significant 
contribution to $I_{\sigma_c\sigma_i}$. The usual characterization 
of the final state with two quantum numbers ($d$-$d$, $s$-$d$, 
etc.) is no longer valid if correlations are included.
The orbital character may change by electron scattering.

Neglecting electron correlations altogether and assuming the matrix 
elements to be constant, Eq.\ (\ref{eq:int}) reduces to the Lander model:
In this case the intensity $I_{\sigma_i}$ is given by the following sum 
of (singlet and triplet) self-convolutions:
\begin{eqnarray}
  I_{\sigma_i} & \propto & 
  \sum_{L_1L_2} \int dE' \:
  \widetilde{\rho}_{L_1\sigma_i}(E') \widetilde{\rho}_{L_2 -\sigma_i}(E-E')
\nonumber \\
  &+& {\sum_{L_1L_2}}' \int dE' \:
  \widetilde{\rho}_{L_1\sigma_i}(E') \widetilde{\rho}_{L_2 \sigma_i}(E-E')
  \: .
\label{eq:lander}
\end{eqnarray}
Here $\widetilde{\rho}_{L\sigma}(E) = (1-f(E)) {\rho}_{L\sigma}(E)$ is the 
unoccupied part of the $L$-resolved and spin-dependent density of 
states (DOS) ${\rho}_{L\sigma}(E)$ where $f(E)$ is the Fermi function.
The prime in the second sum in Eq.\ (\ref{eq:lander}) excludes the term 
$L_1=L_2$, $\sigma_c=\sigma_i$, which is forbidden by the Pauli principle.
Note that lattice symmetries require the on-site ($i=i'$) but off-diagonal 
($L \ne L'$) DOS $\rho_{LL'\sigma}(E) = (-1/\pi) \mbox{Im} 
\langle \langle c_{iL\sigma} ; c_{iL'\sigma}^\dagger \rangle \rangle_E$
to vanish identically. This is a consequence of the choice of the 
(non-orthogonal) basis set $|iL \sigma \rangle$.

To estimate the significance of correlation effects, the two-particle
Green function in Eq.\ (\ref{eq:int}) is calculated for a nine-band 
Hubbard-type model $H=H_{\rm LDA}+H_{\rm int}-\Delta H$ including 
$4s$,$4p$ orbitals and correlated $3d$ orbitals:
\begin{eqnarray}
  H &=& \sum_{ii'LL'\sigma} T_{ii'}^{LL'} \,
  c_{iL\sigma}^\dagger c_{i'L'\sigma} 
\nonumber \\  
   &+& \frac{1}{2} \sum_{i\sigma \sigma'}\sum_{m_1 \ldots m_4} \! 
   U_{m_1 m_2 m_4 m_3} \,
   c_{im_1\sigma}^\dagger
   c_{im_2\sigma'}^\dagger 
   c_{im_3\sigma'} 
   c_{im_4\sigma} 
\nonumber \\  
   &-& \Delta H \: .
\label{eq:hamilton}
\end{eqnarray}
The hopping parameters $T_{ii'}^{LL'}$ of the one-particle term $H_{\rm LDA}$ 
(and also the overlap $S_{ii'}^{LL'}$ parameters) are obtained from a 
Slater-Koster fit to the paramagnetic LDA band structure for Ni. \cite{WPN00}
Opposed to photoemission spectroscopy, this comparatively simple tight-binding 
parameterization appears to be sufficient in the case of APS since the 
two-particle spectrum does not crucially depend on the details of the 
one-particle DOS.

The on-site interaction among the $3d$ electrons is described by the second
term $H_{\rm int}$. The Coulomb matrix depends on four orbital indices, 
$U_{m_1m_2m_3m_4}$, referring to the MTO's for $l=2$.
Using atomic symmetries the interaction parameters can essentially
be expressed in terms of two parameters $U$ and $J$. \cite{WPN00}
Interactions involving delocalized $s$ and $p$ states are assumed to be 
sufficiently accounted for within the LDA.
For the $d$ states, however, there is the well-known double-counting problem: 
The interactions are counted twice, once in $H_{\rm LDA}$ (i.\ e.\ on a 
mean-field level) and once more in $H_{\rm int}$.
To avoid this double counting, a third term $\Delta H$ has been introduced 
by which the Hartree-Fock part of $H_{\rm int}$ is subtracted.
\cite{WPN00,SAS92}

The Hamiltonian $H$ constitutes an involved many-body problem. 
Due to the low density of $3d$ holes in the case of Ni, however, it appears 
to be reasonable to employ the so-called ladder approximation \cite{Cin79} 
which extrapolates from the exact (Cini-Sawatzky) solution for the limit of 
the completely filled band. \cite{Cin77,Saw77}
For finite hole densities the ladder approximation gives the two-particle 
Green function 
$\langle \langle c\, c \, ; c^\dagger c^\dagger \rangle \rangle$ 
as a functional of the one-particle Green function (``direct correlations''). 
The one-particle Green function of the type 
$\langle \langle c\, ; c^\dagger \rangle \rangle$ 
corresponds to the (inverse) photoemission spectrum which itself
is renormalized by $H_{\rm int}$ (``indirect correlations''). 
It is calculated self-consistently within second-order perturbation theory 
(SOPT) around the Hartree-Fock solution. \cite{WPN00}
For a moderate $U \sim \Delta$ and a low hole density, a perturbational 
approach can be justified. \cite{SAS92} A resummation of higher-order diagrams 
is important to describe bound states (``Ni 6 eV satellite'') \cite{Lie81} 
which, however, are relevant for AES only.
The numerical values for the direct and exchange interaction, $U=2.47$~eV and 
$J=0.5$~eV, are taken from Ref.\ \onlinecite{WPN00} where they have been 
fitted to the ground-state magnetic properties of Ni. Since
spin-wave excitations are neglected in the approach, 
the calculated Curie temperature 
$T_{\rm C}=1655$~K is about a factor 2.6 too high compared with the experiment. 
Using reduced temperatures 
$T/T_{\rm C}$, however, the temperature trend of the magnetization 
is well reproduced. \cite{WPN00}

\section{Results and discussion}

The solid lines in Fig.\ \ref{fig:spectrum} (left) show the spectra as 
calculated from Eq.\ (\ref{eq:int}) using the ladder approximation.
To account for apparatus broadening, the results have
been convoluted with a Gaussian of width $\sigma = 0.6$~eV (see Ref.\ 
\onlinecite{EP97}). 
The calculated data are shifted by 852.3~eV such that onset 
of the unbroadened spectrum for $T/T_{\rm C}=0.16$ coincides with the 
maximum of $L_{\rm III}$ emission in the experiment (dotted line).

What are the signatures of electron correlations?
To estimate first the effect of the {\em direct} interaction between 
the two additional final-state electrons, Fig.\ \ref{fig:spectrum} (right) 
also displays
the results of the self-convolution model for comparison (still including 
matrix elements as well as the fully interacting one-particle DOS).
Fig.\ \ref{fig:spectrum} and also a more detailed inspection show that the 
secondary peak at $E \approx 859$~eV is not affected by correlations at all.
This is consistent with observed temperature independence of the peak
and with the fact that the DOS has mainly $s$-$p$ character at the
discontinuity deriving from the $L_7$ critical point.
The maximum of the secondary peak is therefore used as a reference 
to normalize the measured spectra for each $T$.
Looking at the results of the ladder approximation, the overall agreement 
with the measurements is rather satisfying.
Except for the lowest temperature the intensity, the spin splitting and 
the spin asymmetry of the main peak are well reproduced and, 
consistent with the experiment, a negligibly small intensity asymmetry 
and spin splitting is predicted for the secondary peak.
Switching off the direct correlations (Fig.\ \ref{fig:spectrum}, right)
results in a strong overestimation of the main peak structure. 
Within the Cini-Sawatzky theory this has a plausible qualitative 
explanation:
For low hole density the main effect of the direct correlations is 
known to transfer spectral weight to lower energies inaccessible to APS. 
This weight shows up again in the Auger spectrum. 
Recall that in fact a considerable weight transfer is seen in AES 
\cite{Ram91,BFHLS83} and that hypothetically for $U\mapsto \infty$ all weight 
would be taken by a satellite split off at the lower boundary of the 
Auger spectrum. \cite{Cin77,Saw77}

The {\em indirect} correlations manifest themselves as a renormalization 
of the one-particle DOS.
Thus, in first place they are responsible for the correct temperature 
dependence of the intensity asymmetry of the main peak in the AP spectrum.
In particular, the indirect correlations result in a narrowing of the 
DOS, which near the Fermi energy $E_{\rm F}$ is given by the 
(almost spin independent)
quasi-particle weight $z \approx 0.88 < 1$, and in an intrinsic lifetime 
broadening $\Gamma \propto (E-E_{\rm F})^2, T^2$. 
For $T=0$ the latter turns out to be smaller than $\Gamma \approx 0.01$~eV
for $E-E_{\rm F} < 1$~eV. \cite{WPN00}
The discrepancy between experiment and theory for $T/T_{\rm C} = 0.16$ 
is possibly due to an underestimation of $\Gamma$ within the SOPT:
With decreasing $T$ the increase of the exchange splitting leads to the 
appearance of a strong peak just above $E_{\rm F}$ in the minority DOS. 
A larger lifetime broadening of this peak would imply a less pronounced feature
in the differential AP spectrum -- mainly in the minority channel.

\begin{figure}[t]
\begin{center}
\epsfig{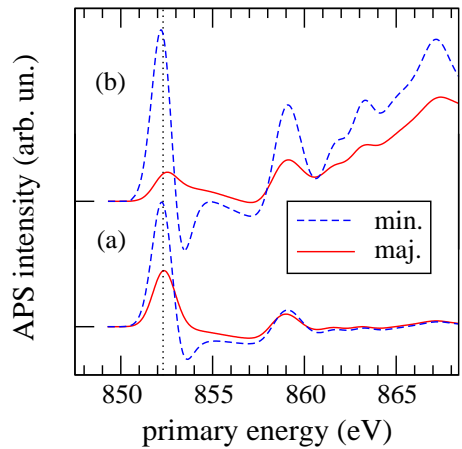}

\caption{
Ni AP spectrum for $T=0$.
(a) full theory. 
(b) as (a) but matrix elements taken to be constant (see text).
\label{fig:tme}
} 
\end{center}
\end{figure}

Setting $M_{L_1L_2}^{\sigma_c\sigma_i}({\bf k}_\|,E) = \pm 1 = \mbox{const}$ 
for $L_1 \ge L_2$ or $L_1 < L_2$, respectively, (see below) and comparing 
with the results of the full theory, demonstrates the importance of the 
transition-matrix elements.
Their energy dependence (via $| {\bf k}_\| E \sigma_i \rangle$) is weak over 
a few eV at energies of the order of keV and cannot explain the difference 
between (a) and (b) in Fig.\ \ref{fig:tme}.
The main difference is rather a consequence of the fact that the radial $2p$ 
core wave function has a stronger overlap with the (more localized) $3d$ as 
compared to the (more delocalized) $4s/4p$ radial wave functions.
This implies a suppression of the $s$-$p$ contributions to the orbital sum 
in Eq.\ (\ref{eq:int}) as already found by Ebert and Popescu. \cite{EP97}
The features above $E=860$~eV originate from additional discontinuities of 
the $s$-$p$-like DOS (as for the peak at $E \approx 859$~eV).

For $T<T_{\rm C}$ the spin asymmetry of the spectrum is mainly due to the 
spin dependence of the Green function in Eq.\ (\ref{eq:int}). 
If the calculation of the matrix elements (\ref{eq:tme}) starts from the 
{\em spin-polarized} L(S)DA potential, an additional spin asymmetry is 
observed resulting from the spin dependence of the states 
in Eq.\ (\ref{eq:tme}). 
This, however, is small and has practically no influence on the results.
On the other hand, Fig.\ \ref{fig:tme} shows a strong suppression of the 
intensity asymmetry at high energies when taking matrix elements into account.
This effect is controlled by the symmetry of the matrix 
$M_{L_1L_2} \equiv M_{L_1L_2}^{\sigma_c\sigma_i}({\bf k}_\|,E)$.
In the antisymmetric case, $M_{L_1L_2} = - M_{L_2L_1}$ ($L_1 \ne L_2$), 
there is a maximum spin asymmetry (Fig.\ \ref{fig:tme}) while, even for 
a ferromagnet, there is no spin asymmetry at all for the symmetric case.
The results of the full calculation along Eq.\ (\ref{eq:tme}) are neither fully
symmetric nor antisymmetric with respect to $L_1$, $L_2$.

\section{Summary} 

In conclusion, the AP line shape of a typical ferromagnetic $3d$ transition 
metal results from a rather complex interplay of different factors.
The present study has shown that a quantitative theoretical analysis of 
the temperature- and spin-dependent spectrum must be based on three 
ingredients at least: 

(i) ``indirect'' correlations: 
A realistic Hubbard-type model including $s$-$p$ like states is a proper 
starting point to describe the magnetism and the temperature-dependent 
renormalization of the one-particle DOS.

(ii) ``direct'' correlations:
While $s$-$p$ derived features at higher energies appear to be sensitive 
to the geometrical structure only, the main peak is strongly affected by 
the direct interaction between the two additional final-state electrons. 
Consistent with the Cini-Sawatzky model, the dominant effect is a considerable 
spectral-weight transfer to energies below the threshold.

(iii) matrix elements:
Spin-resolved APS cannot be described theoretically without calculating
orbital-dependent matrix elements. 
The spin asymmetry is mainly determined by their transformation behavior 
under exchange of the orbital quantum numbers $L_1 \leftrightarrow L_2$.

An open question concerns the importance of core-hole effects in APS. 
Future work shall be concerned with the scattering at the core-hole potential
in the final state and shall include the edge effects \cite{ND69} known from
the studies of simple metals.
A generalized ladder approximation including the valence-core interaction 
has been proposed in Ref.\ \onlinecite{PBB94}.
Previous work \cite{PBNB93,PBB94,PBBN95} has shown that one should not 
expect the effects due the scattering at the core-hole potential to be strong
for the case of Ni.
However, core-hole effects will become more important for systems with a 
smaller $3d$ occupancy.
For Co and Fe one also expects stronger effects of $d$-$d$ correlations.
This work has shown that these $d$-$d$ correlations cannot be neglected
even for a system with low $d$-hole density such as Ni and that 
de-convolution techniques to extract the unoccupied local 
DOS have to be questioned seriously.

\section*{Acknowlegdements}

Support of this work by the DFG (NO~158/5) and by the BMBF (no.~05605MPA0)
is greatfully acknowledged.

\end{document}